# A Survey of Energy Efficiency in SDN: Software-based Methods and Optimization Models

Beakal Gizachew Assefa, Öznur Özkasap*

*Department of Computer Engineering, Koç University, Istanbul, Turkey*

ABSTRACT

Software Defined Networking (SDN) paradigm has the benefits of programmable network elements by separating the control and the forwarding planes, efficiency through optimized routing and flexibility in network management. As the energy costs contribute largely to the overall costs in networks, energy efficiency has become a significant design requirement for modern networking mechanisms. However, designing energy efficient solutions is non-trivial since they need to tackle the trade-off between energy efficiency and network performance. In this article, we address the energy efficiency capabilities that can be utilized in the emerging SDN. We provide a comprehensive and novel classification of software-based energy efficient solutions into subcategories of traffic aware, end system aware and rule placement. We propose general optimization models for each subcategory, and present the objective function, the parameters and constraints to be considered in each model. Detailed information on the characteristics of state-of-the-art methods, their advantages, drawbacks are provided. Hardware-based solutions used to enhance the efficiency of switches are also described. Furthermore, we discuss the open issues and future research directions in the area of energy efficiency in SDN.

## 1. Introduction

Software Defined Networking (SDN) paradigm has been attracting an increasing research interest, with its key concepts of control plane and data (forwarding) plane separation and logically centralized control by means of programmable network devices (Nunes et al., 2014; Kreutz et al., 2015).

SDN has been deployed in a diverse set of platforms ranging from institutional networks to data center networks. It promises several advantages such as flexibility without sacrificing forwarding performance, high efficiency through optimized routing, ease of implementation and administration, and cost reduction. The energy consumption constitutes a significant portion of overall information and communication technology costs (Chiaraviglio et al., 2012; Lambert et al., 2008; Assefa and Ozkasap, 2015). Energy constitutes more than 10% of OPEX (operating expenses) of an ICT service provider. SMARTer 2020 report predicts that the electricity cost of the cloud data centers will increase by 63% in 2020 (GreenPeace, 2014). Several survey studies have been conducted on reducing energy costs in different network settings such as P2P systems (Brienza et al., 2016), Network Function Virtualization (NFV) (Mijumbi et al., 2016), cloud data centers (Hammadi and Mhamdi, 2014), and wireless sensor networks (Soua and Minet, 2011). However, to the best of our knowledge, an extensive survey and classification of recent developments in energy efficiency for SDN along with optimization models for each group of solutions is not available.

We address energy optimization that can be applied at various levels of the SDN architecture, or SDN itself that can be used as a means of energy saving. Fig. 1 shows the categories of energy saving approaches in SDN. Energy saving in SDN can be addressed algorithmically or through hardware-based improvements. Software-based solutions are applied on the controller. The three energy saving capabilities that can be addressed algorithmically are traffic aware, end system aware and rule placement.

Traffic awareness is the capability to make the energy consumption proportional to the traffic volume. End system awareness on the other hand uses SDN to save energy by capitalizing on virtual machine placements and migrations. Most energy aware routing algorithms do not assume the limited rule space inside switches. Rule placement techniques address the energy saving from the meaning of the rules, and compressing them in way they save space.

* Corresponding author.
  *E-mail address:* oozkasap@ku.edu.tr (Ö. Özkasap).

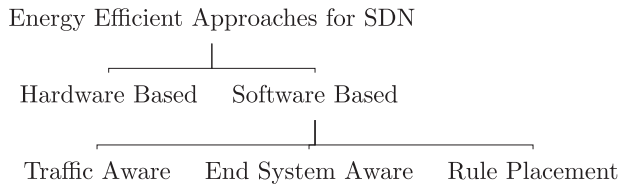

**Fig. 1.** Energy saving capabilities in software defined networking.

Table 1 shows the list of survey works on the topic of SDN and their respective objectives. The columns of the table indicate the theme and scope, and the symbol + tells if the topic is covered in the survey. The themes of the related survey works range from general introduction to SDN, architectures, and applications (Nunes et al., 2014; Xia et al., 2015; Kreutz et al., 2015), to security (Rawat and Reddy, 2017; Dargahi et al., 2017; Zhang et al., 2018), energy efficiency (Assefa and Ozkasap, 2015; Tuysuz et al., 2017), rule placement (Nguyen et al., 2016), load balancing (Neghabi et al., 2018), and traffic engineering (Mendiola et al., 2017). Besides, there exist surveys on the usage of SDN in data center networks, optical networks, wireless networks, transportation networks, edge computing, and a hybrid of traditional networking with SDN (Thyagaturu et al., 2016; Alvizu et al., 2017; Kobo et al., 2017; Ndiaye et al., 2017; Baktir et al., 2017). This survey extends our prior work (Assefa and Ozkasap, 2015) which provided a brief review of energy saving methods in SDN, to the best of our knowledge, for the first time in the literature. There exist other subsequent studies providing survey of energy efficiency of SDN (Rawat and Reddy, 2017; Tuysuz et al., 2017). In contrast to (Rawat and Reddy, 2017), we do not focus on SDN as a tool for security and its trade-off with energy efficiency, rather we focus on extensive discussion of energy efficient solutions. Unlike (Tuysuz et al., 2017), we present the optimization models, the objective functions, constraints, and provide discussion of methods under each group of energy saving capabilities we have identified.

In contrast to prior work, we categorize and evaluate state-of-the-art energy efficient solutions in SDN considering both software-based and hardware-based approaches. Our survey differs from previous studies by focusing on the software-based approaches, where we identify subcategories as traffic aware, end system aware and rule placement. For each subcategory, we present general optimization models and key characteristics of state-of-the-art solutions with their pros and cons, and their comparison with the models. As the main contributions in this article, we:

- address the energy efficiency capabilities that can be utilized in SDN. Energy efficiency can be addressed algorithmically or by hardware design.
- present a taxonomy of software-based energy efficient solutions in SDN into subcategories of traffic aware, end system aware and rule placement.
- propose general optimization models for each subcategory, and for each model we present the objective function, the parameters to be considered and constraints that need to be respected.
- present key characteristics of state-of-the-art solutions for each category, their advantages, drawbacks, and provide comparisons with the general models.
- discuss hardware-based solutions used to enhance the efficiency of a switch, and state open issues and research directions for energy efficiency in SDN.

The remainder of the article is organized as follows. We describe principles of SDN and terminology in Section 2. Section 3 presents traffic aware solutions, optimization model and methods in the literature. The end system aware solutions, multiple objective optimization model and techniques are presented in Section 4. Section 5 presents the rule placement problem in SDN, model for optimizing rule space for energy efficiency and example methods whereas section 6 presents hardware-based energy efficient solutions in SDN. In Section 7, we discuss open issues on energy efficiency in SDN and provide guidelines for future research, followed by concluding remarks in Section 8.

## 2. Background and terminology

### 2.1. Traditional networking vs SDN

In traditional networking, the data plane and the control plane reside in the switch. The control plane populates the rules required for forwarding into the forwarding table. Whereas, the forwarding plane reads the forwarding table to forward packets. The control decisions taken at a given switch determines the next hop in the network that each packet needs to be forwarded. The intelligence in networking which includes security, forwarding, and load balancing at switch level. Since the control plane and the data plane are highly coupled in every networking component, adapting any change in the networks is cumbersome because it demands every component of the network to be configured.

SDN, on the other hand, is a novel paradigm in networking with the key design principle of the control plane and the data plane separation. This design principle, where the data plane resides at the switches and the control plane resides in the controller, provides flexibility network management. The forwarding table is inside the networking components, but the rules are created and pushed to each switch by the controller. The prominent advantage of SDN is that it provides a global view of the network to be controlled and monitored centrally. The switches in SDN only do packet forwarding. The control is done by the logically centralized controller. SDN makes the network programmable, and enables changes done on the controller to propagate to the entire network.

**Table 1**
Comparison of survey works on Software Defined Networks.

| Survey | Theme | General | Performance | Security | Energy | Rule Placement | End System | Formulation | Hardware |
|---|---|---|---|---|---|---|---|---|---|
| (Nunes et al., 2014; Xia et al., 2015; Kreutz et al., 2015) | Overview | + | | | + | | | | |
| Nguyen et al. (2016) | Rule Placement | | | | | + | | | |
| Thyagaturu et al. (2016) | Optical NW | | | | | | | | |
| Alvizu et al. (2017) | Transport NW | | + | | | | | | |
| (Kobo et al., 2017; Ndiaye et al., 2017) | Wireless | + | + | | | | | | |
| Baktir et al. (2017) | Edge Computing | | | | | | | | |
| Rawat and Reddy (2017) | Security & EE | + | | + | + | + | + | | |
| Mendiola et al. (2017) | Traffic Eng. | | + | | + | | | | |
| (Assefa and Ozkasap, 2015; Tuysuz et al., 2017) | Energy | | + | | | + | + | | + |
| (Dargahi et al., 2017; Zhang et al., 2018) | Security | | + | + | | | | | |
| Amin et al. (2018) | Hybrid | | | | | | | | |
| Neghabi et al. (2018) | Load Balancing | | + | | + | | + | | |
| **Our survey** | Energy | | + | | + | + | + | + | + |

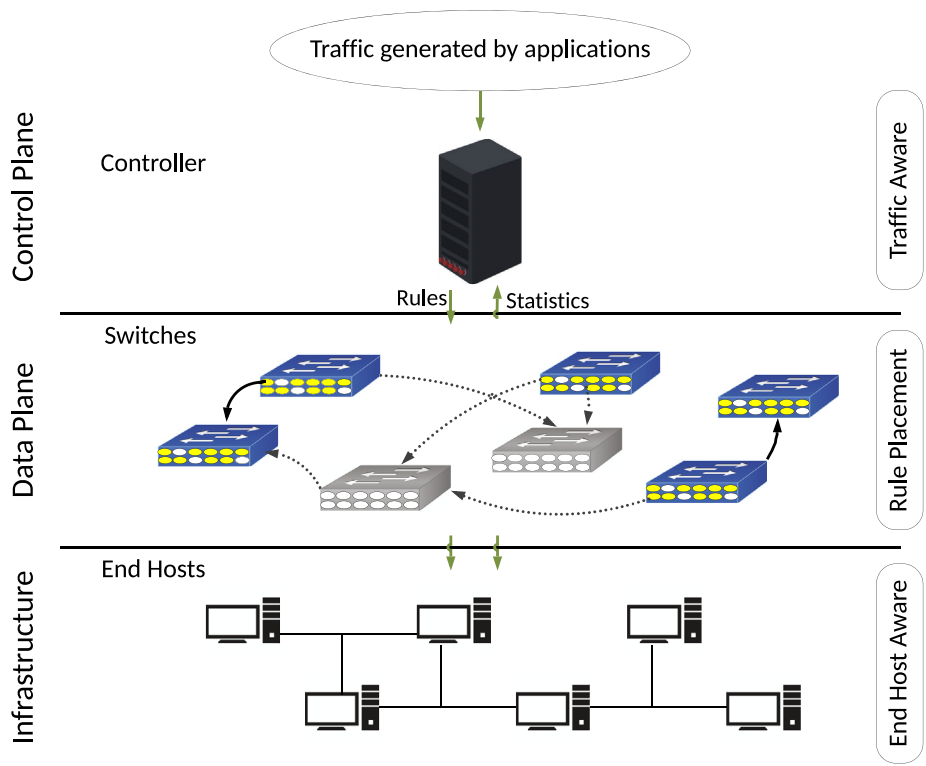

**Fig. 2.** Software defined networking architecture.

The idea behind SDN paradigm, depicted in Fig. 2, is eliminating the tight coupling between control and forwarding functions in traditional network design, and hence the drawbacks of cumbersome network configuration and limited flexibility to changing requirements (Nunes et al., 2014). The management and control of the network is done in the control plane. The controller has the functionality of configuring the forwarding tables (in other words, flow tables) of switches. The set of connected switches constitutes the data plane of the SDN architecture. The sole responsibility of the data plane is to forward packets based on the forwarding rules installed in the flow tables of the switches by the controller. Fig. 2 also illustrates our perspective of energy saving capabilities in the SDN architecture.

### 2.2. Desired properties of software based energy saving capabilities in SDN

Fig. 3 illustrates the categories of software based energy saving approaches in SDN. Software based energy saving approaches are divided in to traffic aware, end system aware and rule placement.

**Traffic aware** energy efficiency approaches are inspired by the fact that network components are often under utilized. The key principle is to turn on or turn off network components (i.e., SDN forwarding switches) based on the traffic load. For instance, when the traffic load is low (e.g., during night times) this approach has the potential to save up to 50% of the total energy consumption (Heller et al., 2010). Typically, an elastic structure is used to represent the network components that can grow and shrink with the dynamic traffic load. The key challenge is to determine which components to turn on and which components to turn off without compromising the required quality of service (QoS) (Heller et al., 2010; Staessens et al., 2011; Canini et al., 2011; Wang et al., 2012; Thanh et al., 2012; Nguyen et al., 2013; Wang et al., 2014a; Vu et al., 2014; Vu et al., 2015; Giroire et al., 2015; Rodrigues et al., 2015; Zhu et al., 2016; Riekstin et al., 2016).

As presented in Fig. 3, desirable properties of a **traffic aware** controller are elasticity, topology awareness, queue engineering, and smart sleep on and off. *Elasticity* is the ability to dynamically enlarge or shrink the number of network components used in response to traffic. *Topology awareness* provides an extra benefit of using formulations and solvers that can be tailored to any specific topology. The hierarchically organized fat-tree is the widely used topology in data centers. A prior knowledge to how the components are organized and their capacity allows us to use alternative routes by avoiding energy critical paths. *Queue Engineering* techniques applied to each port for the packet arrivals give additional port level traffic monitoring capability. *Smart sleep and off* is the ability to turn on/off ports of switches, links, or the entire switch to save energy in response to traffic.

**End system aware** energy saving solutions use the practice of turning off underutilized physical servers and running their tasks on a fewer number of servers in SDN based data centers (Shirayanagi et al., 2012; Jiang et al., 2012; Jin et al., 2013; Wang et al., 2014b; Zheng et al., 2014; Chen et al., 2015; Dalvandi et al., 2015; Ortigoza et al., 2016). Specifically in data centers, the SDN model is used to form an overlay connecting virtual machines. The overlay is a list of SDN enabled switches and links. The actual connection is of course with the physical machines which host the virtual machines.

Desirable properties of an end system aware solution are depicted in Fig. 3. *Server consolidation* is the technique of minimizing the number of physical machines that are active by placing as many virtual machines as possible to fewer physical machines. *Network optimization* on the other hand minimizes the transport cost of migrating virtual machines from one physical machine to the other one. *Dynamicity* is the ability to apply both server consolidation and network optimization online instead of offline. The need for online migration of virtual machines and network optimization in response to traffic and workload is of great value. *Reusability* is the characteristic of re-booking resources after they are released from prior works. This comes from the fact that some resources can be used for longer period of time as compared to others.

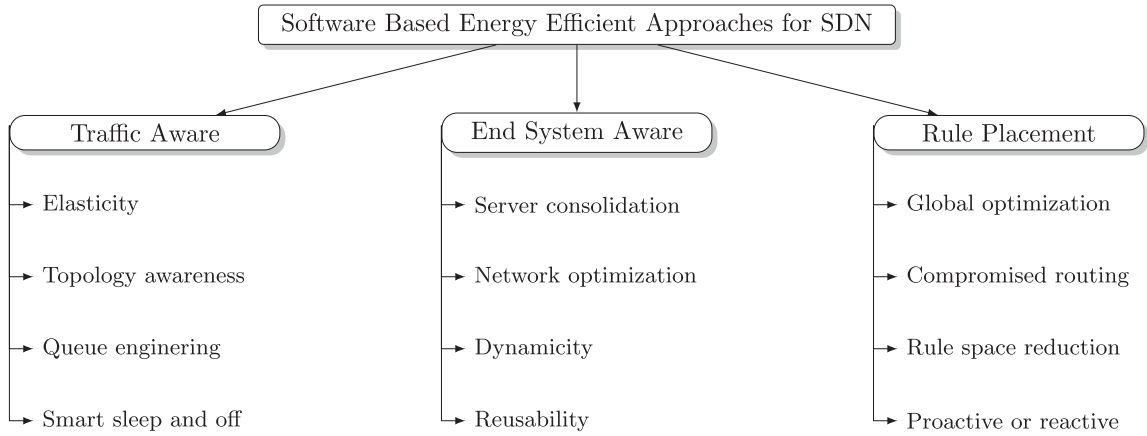

**Fig. 3.** Classification: Software based energy efficiency approaches in SDN.

**Rule placement** techniques focus on how to place the rules in the forwarding switches. Given the network policies and end-point policies, the controller provides a mechanism to convert the high level policies into switch understandable rules. Rule placement is an NP-hard problem and hence requires heuristic based solutions. Although heuristic based approaches do not guarantee optimal solution, they typically offer close to optimal results. And their efficiency depends on the complexity of the constraints (Kang et al., 2013; Kanizo et al., 2013; Giroire et al., 2014, 2016; Nguyen et al., 2015; Rifai et al., 2015). Some of the constraints considered for this problem are the maximum number of rules a switch can hold, the routing policy, and the topology. Under such constraints, rule placement approaches attempt to optimize routing.

Desirable properties of a rule placement algorithm are depicted in Fig. 3. **Global view** is the ability to use entire network information, routing policy, and end-point policy to decide which rules to place on which switch. **Compromised routing** is the ability to find alternative power efficient routes at the cost of performance. However, the trade off between compromised routing performance due to extra longer paths and its impact on QoS should be taken with great care. The set of all rules that need to be installed on the switches constitute the rule space. **Rule space reduction** is the ability to reduce the rule space, hence, minimize the number of active links and or switches. **Proactive/reactive** is the ability to do energy efficient rule placement proactively before the packets arrive to the network, and respond reactively to newly arriving packets.

As far as the software-based solutions are concerned, the controller should be acquainted with the necessary statistical information about the status of the network devices and end systems attached, in order to devise a way for energy efficient networking. A full fledged energy efficient controller needs to consolidate end systems, network device optimization, and also be able to reduce the rule space.

*2.3. Terminology*

Table 2 presents notation used in this paper for formulating the energy efficiency problem. The symbols are listed in the order they are used in the paper. The set of **switches** are denoted by $\mathbb{Z}$ where each switch ($Z_i$) is a forwarding device with the role of forwarding packets based on their flow information. $S_i$ is a binary variable with value 1 if the switch $Z_i$ is turned on, or 0 otherwise. The total power consumption of $Z_i$ is denoted by $CS_i$.

Switches are connected to one another by links. A **link** is active if it is transmitting packets between the two ports of the switches it connects. $L_{ij}$ is a binary variable with value 1 if the link connecting switches $i$ and $j$ is active, or 0 otherwise. The power consumption of a link connecting switches $i$ and $j$ is denoted by $C_{ij}$. $G_i$ and $W_{ij}$ are the maximum number of rules that can be installed in switch $i$ and the bandwidth of the link connecting switches $i$ and $j$, respectively.

A network in SDN is composed of forwarding switches connected by links, which is transmission media for data among end systems. An **end system** is a physical or a virtual machine where services and applications are running on. The set of physical machines and virtual machines are denoted by $\mathbb{P}$ and $\mathbb{V}$, respectively. Each physical machine has a limited set of resources $\mathbb{R}p$. A virtual machine $i$ has set of hardware or software demand $\mathbb{R}v$ to be able to run on a physical machine. $X_{ij}$ is a binary variable with value 1 if virtual machine $j$ is placed on physical machine $i$. $PM_i$ is a binary variable with value 1 if physical machine $i$ is on or 0 otherwise. $\mathbb{R}p$ and $\mathbb{R}v$ are represented as matrix of resources available on physical machines and resource demands of virtual machines, respectively. Vector $\mathbb{R}$ represents the resource types. The matrix $\mathbb{R}p$ and $\mathbb{R}v$ have the dimensions of $|\mathbb{P}|$ by $|\mathbb{R}|$ and $|\mathbb{V}|$ by $|\mathbb{R}|$, respectively. A set of virtual machines is placed on a physical machine if and only if the sum of the resource demands of virtual machines can be met by the available resources on the physical machine. The traffic matrix between virtual machines is denoted by $\mathbb{Q}$, where $q_{ij}$ represents the traffic between virtual machines $i$ and $j$ measured in bps. $b_{ij}$ is the number of switches the traffic between physical machines $i$ and $j$ traverses.

A switch contains a **flow table** or set of **flow tables**. The fields of a flow table may vary from vendor to vendor. OpenFlow 1.0 switch specification presents 12 packet matching fields (header fields), counter, and action. The matching fields have information that can be compared with packets arriving to the switch. Per-flow, per-switch, and per-port counters are maintained in the flow table. The action field is an instruction of what to be done with the corresponding packet matching the header field. According to OpenFlow 1.4 standard, a switch can have multiple tables and matching is done through pipeline process among the tables.

Packets flow through the network starting from their source towards destination. The controller decides the path of the packet. An important issue while managing the network is the level of granularity of the control. Packet level control makes the controller congested since a new rule should be created and pushed for every packet, whereas prefix-based matching reduces the control capability since several packets matching a prefix would be treated the same way. As an intermediate level of granularity, flows are introduced (Gude et al., 2008).

The set of flows is denoted by $\mathbb{F}$ where a flow is defined as a set of packets with the same source and destination addresses where the packets pass through the same route to reach the destination. $H_{in}$ and $H_{eg}$ are the set of ingress hosts and egress hosts, respectively. Ingress hosts are hosts where flows start from and egress hosts are hosts which are destinations of flows. $\lambda_f$ denotes the packet rate of flow $f$. $G_i$ is the maximum number of rules that can be installed in switch $Z_i$, respectively.

**Table 2**
Parameters and descriptions.

| Symbol | Description |
| --- | --- |
| $\mathbb{Z}$ | Set of switches in the network where $Z_i \in \mathbb{Z}$ represents switch $i$ |
| $\mathbb{E}$ | Edge or links where $e_{ij} \in \mathbb{E}$ represents the link between switches $Z_i$ and $Z_j$ |
| $W_{ij}$ | Bandwidth of the link connecting switches $Z_i$ and $Z_j$ |
| $S_i$ | $= \begin{cases} 1, & \text{if switch } Z_i \text{ is active,} \\ 0, & \text{otherwise.} \end{cases}$ |
| $CS_i$ | Power consumption of switch $Z_i$ |
| $C_{ij}$ | Power consumption of the link connecting switches $Z_i$ and $Z_j$ |
| $\mathbb{F}$ | Set of flows where $f \in \mathbb{F} = (sr, ds, \lambda_f)$ |
| $f = (sr, ds, \lambda_f)$ | A flow f with source, destination and packet rate |
| $F_{ij}$ | $= \begin{cases} 1, & \text{if flow } f \text{ passes through edge } e_{ij}, \\ 0, & \text{otherwise.} \end{cases}$ |
| $\mathbb{P}$ | Set of physical machines |
| $\mathbb{V}$ | Set of virtual machines |
| $\mathbb{R}$ | Vector of resource types |
| $\mathbb{R}p$ | Set of resources available on physical machines represented as a matrix |
| $\mathbb{R}v$ | Set of resource demands of virtual machines represented as a matrix |
| $X_{ij}$ | $= \begin{cases} 1, & \text{if virtual machine } j \text{ is placed on physical machine } i, \\ 0, & \text{otherwise.} \end{cases}$ |
| $PM_i$ | $= \begin{cases} 1, & \text{if physical machine } i \text{ is on,} \\ 0, & \text{otherwise.} \end{cases}$ |
| $P_i^r$ | Amount of resource type r available on physical machine $i$ |
| $V_j^r$ | Amount of resource type r required by virtual machine $j$ |
| $\mathbb{Q}$ | Traffic matrix between virtual machines |
| $q_{ij}$ | Amount of traffic between virtual machines $i$ and $j$ |
| $b_{ij}$ | Number of switches that the traffic between physical machines $i$ and $j$ traverses |
| $H_{in}$ | Set of ingress hosts |
| $H_{eg}$ | Set of egress hosts |
| $\mathbb{A}$ | Allocation matrix, $a_{if} = \begin{cases} 1, & \text{if rule representing flow } f \text{ is installed on switch } Z_i, \\ 0, & \text{otherwise.} \end{cases}$ |
| $L_{ij}$ | $= \begin{cases} 1, & \text{if edge } e_{ij} \text{ active,} \\ 0, & \text{otherwise.} \end{cases}$ |
| $G_i$ | The maximum number of rules switch $Z_i$ can store |

**Rule** in SDN is a predicate that describes a flow. An example of a rule predicate is destination IP = 198.127.***.***. If a packet that has the destination IP address that starts with 198.127, its corresponding action will be taken. The controller populates the flow tables by pushing the rules to each switch.

The rules are stored with priority which is the matching precedence of the flow entry. A rule which all entries are wildcards is given the least priority value. For each packet arriving a switch, its flow information is extracted from the packet header and then looked up in the flow table. Then, the highest priority entry matching the packet is selected. $\mathbb{A}$ is an allocation matrix between flows and switches where $a_{if}$ is a binary variable with value 1 if rule representing flow $f$ is installed on switch $Z_i$ or 0 otherwise.

An **action** defines the task to be performed for the packet that matches a rule. An action for a packet can be forward to the indicated port, drop the packet, default port to forward to the controller, modification of the packet header or any other action depending on the type of protocol used. The action associated with the first rule that matches the packet will be executed. If a packet does not match any rule in the flow table, then it is forwarded to the controller. After a set of rules is installed for a packet, if another packet comes with the same header description, there is no need to push new rules, rather it uses the existing rules.

### 3. Traffic aware solutions

#### 3.1. Overview

Traffic in networking is defined as the amount of packets transmitted through a network at a given point in time. Traffic management involves dynamically analysing, predicting and regulating the behaviour of the network devices to optimize the performance and QoS requirements. Examples based on usage of traffic are load balancing, performance optimization, security, access control, bandwidth management, and energy usage (Akyildiz et al., 2014; Pistirica et al., 2015).

The logically centralized controller in SDN has a global view of the network. The controller periodically gathers statistical information about the traffic, the status of the switches, the status of the links, the status of the end systems, and the topology of the network. Programmability of a network, a powerful property of SDN, enables the controller adapt to the changing environment. A change in the network environment can be caused by addition of a new host, failure of a network component, or modification of network policy. Network components work with full capacity in all times regardless of the traffic demand. The power consumption of the network is not proportional to the volume of traffic (Heller et al., 2010). With traffic aware energy efficiency approaches, energy consumption can be reduced by turning off some forwarding switches during low traffic load, or putting CPUs or ports at sleep mode. The solutions in this group have the potential to significantly improve energy efficiency in SDN. For data centers, traffic aware approach achieves power savings of up to 50% during low load periods (Heller et al., 2010).

#### 3.2. Traffic aware model

Based on the review of various models used to capture traffic proportional energy consumption, we propose a general optimization model that identifies energy capabilities in SDN from traffic point of view, where the major energy saving components are considered as the links and the switches. The model jointly minimizes the number of switches and the number of links used to accommodate the given traffic. It is tailored to include additional parameters and constraints. For simplicity, the parameters of the optimization model refer to snapshot of the network state. Given the flows which represent the traffic, the problem is defined as allocating links and switches for each flow while minimizing the total number of active links and switches. Inspired by the approaches of (Heller et al., 2010; Giroire et al., 2015; Rui et al., 2015; Zhu et al., 2016), we use multi-commodity flow problem formulation. Each flow is treated as a commodity with source, destination, and flow rate. Our model is inspired by the formulation of (Heller et al., 2010) and the objective of the model is to minimize the power consumption of the links and switches used to handle the flows. The other approaches in this category consider additional parameters like time, bandwidth awareness, and redundant path elimination.

In traffic aware model, the network is represented as an undirected weighted graph $\mathbb{G} = (\mathbb{Z}, \mathbb{E})$ where $\mathbb{Z}$ is the set of switches and $Z_i \in \mathbb{Z}$ represents switch $i$ and $e_{ij} \in \mathbb{E}$ represents that there is link between switches $Z_i$ and $Z_j$. The weight $W_{ij}$ corresponds to the bandwidth of the link connecting switches $Z_i$ and $Z_j$. Let binary variable $S_i$ denote the status of switch $Z_i$ such that

$$S_i = \begin{cases} 1, & \text{if switch } Z_i \text{ is active} \\ 0, & \text{otherwise} \end{cases}$$

$CS_i$ and $C_{ij}$ are power consumption of switch $Z_i$ and the link $e_{ij}$ measured in watt.

Traffic in the network is represented by set of flows $\mathbb{F}$ where $f \in \mathbb{F}$ is defined as $f = (sr, ds, \lambda_f)$. $sr$ and $ds \in \mathbb{Z}$ are the source and destination switches and $\lambda_f$ is the rate of flow $f$ measured in bytes per second.

$$F_{ij} = \begin{cases} 1, & \text{if flow } f \text{ passes through edge } e_{ij} \\ 0, & \text{otherwise} \end{cases}$$

The multi-objective function (Eq. (1)) minimizes the sum of the energy consumption of the switches and the links. The first item in the objective function, the sum of $F_{ij} * C_{ij}$, refers to the total energy consumption incurred by all flows using edge $e_{ij}$. The second item is the total power consumption of all active switches in the network. The objective function jointly minimizes the sums subject to the constraints described next.

$$\text{minimize } (\sum_{\forall f} \sum_{\forall e_{ij}} F_{ij} * C_{ij} + \sum_{\forall S_i} S_i * CS_i) \quad (1)$$

$$\text{subject to } \sum_{\forall f} F_{ij} * \lambda_f \leq W_{ij}, \forall e_{ij} \quad (2)$$

$$\sum_{\forall f} F_{ai} = \sum_{\forall f} F_{ib}, Z_i \neq sr, ds \in f_{sr, ds, \lambda_f} \quad (3)$$

$$F_{mj} = F_{in}, Z_m = sr, Z_n = ds, \forall e_{mj}, \exists e_{in} \quad (4)$$

$$F_{ij} \leq S_j, \forall Z_j \in \mathbb{Z} \quad (5)$$

$$F_{ij} \leq S_i, \forall Z_i \in \mathbb{Z} \quad (6)$$

$$S_i \leq \sum_{\forall f} [F_{ij} + F_{ji}], \forall Z_i \in \mathbb{Z} \quad (7)$$

The constraint given in Eq. (2) states that the total rate of flows between two switches should not exceed the link capacity. Constraint specified in Eq. (3) states that the number of flows entering and leaving for switches which are neither destination nor sources of a flow should be equal. Constraint on Eq. (4) assures a flow entering from source switch should reach the destination switch. The constraints on Eqs. (5)–(7) maintain the correlation between switches and links using the switch state variable and flow-link variables. While constraints 5 and 6 state that no flow should use a link connected to an inactive switch, constraint 7 states that if no flow is passing through the links connected to a given switch, then the switch is switched off. Given the formulation and the parameters of the model, the output of the optimizer is list of links and active switches for each flow $f \in \mathbb{F}$.

### 3.3. Methods

Table 3 shows the summary of traffic aware solutions for energy efficient SDN. The experiment column presents how the proposed solution is tested. The testing environments are testbed, network simulation tools (ns-2), and emulation software (mininet, OMNeT++). The traffic traces used for the test are real world traces (R), artificially generated (A), or both (R&A). The topologies used in the evaluations are fat-tree, BCubic, CERNET, German, SDNLib generated. Controller column indicates the name of the controller used in the experiment (NOX, POX, FL-Floodlight, RY-RYU, and OD-OpenDaylight). The objective of the optimization is to minimize the number of active switches, links/ports or both. QueueEng column presents if a given approach uses switch level queue engineering techniques. The symbol '-' means unknown for Traffic, Topology, and Controller columns, and means does not apply in case of Switch, Link/port and QueueEng columns.

**Table 3**
Traffic aware techniques for energy efficiency.

| Approach | Experiment | | | | Objective | | QueueEng |
|---|---|---|---|---|---|---|---|
| | Environment | Traffic | Topology | Controller | Switch | Link/Port | |
| ElasticTree (Heller et al., 2010) | Testbed | R | Fat-tree | NOX | – | ✓ | – |
| Carrier Grade (Staessens et al., 2011) | Testbed | R | German | NOX | – | ✓ | – |
| REsPoNse (Canini et al., 2011) | NS-2 | R | – | – | ✓ | – | – |
| CARPO (Wang et al., 2012) | Testbed | R | Fat-tree | – | – | ✓ | – |
| EnableOpenflow (Thanh et al., 2012) | Mininet | R & A | Fat-tree | NOX | – | ✓ | – |
| RA-TAH (Nguyen et al., 2013) | Testbed | R & A | Fat-tree | POX | – | ✓ | ✓ |
| Re-Routing (Wang et al., 2014a) | Testbed | R & A | CERNET | – | – | – | ✓ |
| Dynamic TA (Markiewicz et al., 2014) | Testbed | R & A | Mesh | – | ✓ | ✓ | – |
| TE based (Vu et al., 2014) | Testbed | R & A | CERNET | – | ✓ | ✓ | – |
| NetFPGA QE (Vu et al., 2015) | Testbed | – | – | – | ✓ | – | ✓ |
| GreenRE (Giroire et al., 2015) | Testbed | R & A | SDNLib | – | ✓ | ✓ | – |
| Bandwidth-aware (Rui et al., 2015) | OMNeT++ | R | Bcubic & Fat-tree | – | – | – | ✓ |
| GreenSDN (Rodrigues et al., 2015) | Mininet | R | Bcubic & Fat-tree | POX | – | ✓ | ✓ |
| Flow Schedule (Guo et al., 2016) | NS-3 | A | Fat-tree | – | ✓ | ✓ | – |
| OpenNaas (Zhu et al., 2016) | Testbed | – | – | RY, FL, OD | ✓ | ✓ | ✓ |
| Orchestrate (Riekstin et al., 2016) | Mininet | R & A | – | POX | – | ✓ | ✓ |
| Resource-aware (Rahnamay-Naeini et al., 2016) | Testbed | A | Fat-tree | – | ✓ | – | – |
| Multiple Contoller (Fernandez-Fernandez et al., 2016) | Testbed | R | SNDLib | – | – | ✓ | – |
| 5G (Fernández-Fernández et al., 2017) | Testbed | R | SNDLib | – | – | ✓ | – |
| Utility-based (Assefa and Ozkasap, 2017) | Mininet | R | GEANT | POX | – | ✓ | – |
| FLOWP (Luo et al., 2017) | OPNET | A | Fat-tree | – | ✓ | – | – |
| TA framework (Assefa and Ozkasap, 2018) | Mininet | R | SNDLib | POX | ✓ | ✓ | – |
| SDN/Ethernet (Maaloul et al., 2018) | Testbed | R | SNDLib | – | ✓ | ✓ | – |
| MTSDPFPR (Ba et al., 2018) | Mininet | R | GEANT | Floodlight | – | ✓ | – |

As shown in Table 3, most of the traffic aware energy efficient methods focus on minimizing the number of active links primarily and switches secondly or both. Testing energy efficient methods in real time on an actual network equipment is not timely and economically feasible. That is the main reason why researchers resort to create small-scale testbeds or use simulation/emulations environments. Queue engineering is the least used technique and it mostly complements the active link/switch minimizing methods. The network topology and traces provided on SNDLib are widely used data sets in measuring energy efficiency in SDN.

ElasticTree is a power management solution for data center networks which is implemented on a testbed consisting of OpenFlow switches (Heller et al., 2010). The idea is to turn off links and switches based on the amount of traffic load. Therefore, energy consumption of the network is made proportional to the dynamically changing traffic. It consists of three optimizers: formal model, greedy bin-packing, and a topology-aware heuristic. Each optimizer takes network topology (a graph), routing constraints, power model (flat), and traffic matrix as input, and outputs subset of links and flow routes.

The formal model formulates the power saving problem by specifying objective function and constraints. The objective function minimizes the sum of the total number of switches turned on and the number of links. The advantage of the formal model is that it guarantees a solution within some configurable optimum; however, the model only scales up to 1000 hosts.

The greedy bin-packing optimizer evaluates possible paths and chooses in left to right order manner, that is, the leftmost path is chosen first then the selection proceeds till the rightmost path. The optimizer improves the scalability of the formal model. This approach suffers the same problem as any of the heuristic techniques. However, solutions can be computed incrementally and can support on-line usage.

The topology-aware heuristic optimizer, on the other hand, splits the flow and finds the link subset easily. It is computationally efficient, since it takes advantage of a fat tree structure and takes only port counters to compute link subset. This approach uses Integer Programming (IP) to formalize the optimization problem. The drawback is degradation of performance because of turning on and turning off components. The model used in ElasticTree contains more constraints, including fat tree topology constraints, than the general model. Experimental results show that ElasticTree achieves energy savings of up to 50%. One drawback is that it does not consider the correlation among the flows.

In the technique of Carrier Grade (Staessens et al., 2011), the focus is the energy efficiency and resilience characteristics of carrier grade networks. Related to energy efficiency, it is demonstrated that OpenFlow can reduce network wide network energy consumption and improve scalability. MLTE is implemented together with energy saving mechanisms such as controlled adaptive line rates at the switches (Puype et al., 2011). For the resilience, it is shown that OpenFlow can handle failures at the switches and the controller, and perform recovery with flow restoration. Similar to ElasticTree, energy savings of up to 50% are achieved.

REsPoNse is a framework that allows network operators to automatically identify energy-critical paths (Canini et al., 2011). It investigates the possibility to pre-compute a few energy-critical paths that, when used in an energy-aware fashion, can continuously produce close-to-optimal energy savings over long periods of time. REsPoNse identifies energy-critical paths by analyzing the traffic matrices, installs them into a small number of routing tables (called always-on, on-demand, and fail-over), and uses a simple, scalable online traffic engineering mechanism to deactivate and activate network elements on demand. The network operators can use REsPoNse to overcome power delivery limits by provisioning power and cooling of their network equipment for the typical, low to medium level of traffic. An additional formulation added to our general model is identifying the energy critical paths and modeling the links power consumption separately and achieves an energy saving of 40%.

CoRrelation-aware Power Optimization (CARPO) algorithm dynamically consolidates traffic flows onto a small set of links and switches in a data center network, and aims at switching off idle network components to reduce energy consumption (Wang et al., 2012). It consolidates traffic flows based on correlation analysis among flows. Another important feature of CARPO is to integrate correlation-aware traffic consolidation with link rate adaptation for maximized energy savings. The integration is formulated as an optimal flow assignment problem. A near-optimal solution is first computed using integer programming to determine consolidation and the data rate of each link in the data center network. A heuristic algorithm is used to find a consolidation and rate configuration solution with acceptable runtime overheads. The heuristic reduces the computation complexity. In addition to our general model, CARPO introduces parameters to represent correlation among flows. This effort extends the ElasticTree work to be flow similarity aware also achieves 50% efficiency.

A test platform for measuring and analyzing the energy consumption of OpenFlow based data centers is presented in (Thanh et al., 2012). The approach used is both traffic aware and energy aware. The three energy saving capabilities covered are switch, link and port. Considering traffic information, the number of active links and switches is determined. The Adaptive Link Rate (ALR) technique and a routing algorithm similar to ElasticTree are used to determine a route for each flow. The power model of NetFPGA-based OpenFlow switch and used NOX controller are the building blocks of the test platform. Experimental results show that this approach saves 35% energy.

ALR based techniques aim at reducing the energy consumption of a link by scaling the rate of the link proportional to the link utility. Rate Adaptive Topology-Aware Heuristic (RA-TAH) solution utilizes both smart sleeping and power scaling of links to improve energy efficiency of data center networks (Nguyen et al., 2013). A comprehensive survey on ALR techniques can be found in (Bilal et al., 2013). The work is an extension of (Heller et al., 2010). This combined mechanism was deployed in a data center using Fat-Tree topology. The bounds on energy savings in low and high traffic utilization cases were analyzed. Analytical results show that the combined algorithm reduces energy consumption remarkably as compared to the conventional methods in case of high traffic with up to 48% energy saving. Unlike the general model we proposed, this technique only focuses on link utilization rather the switches.

A global power management algorithm based on routing traffic via alternative paths to balance the load of links is presented in (Wang et al., 2014a). A binary integer programming model is used for formulating the energy consumption of integrated chassis and line-cards. The chassis and the line-cards are put to sleep based on link utilization and delay of packets. Less utility of links and large amount of delay of packets give information that either the chassis or the line care or both can be put to sleep. A greedy heuristic algorithm is proposed. Experimental results with scenarios of synthetic topology and real life topology CERNET demonstrate the reduction of power consumption by up to 67%.

Motivated by the fact that energy consumption of routers remains constant even if the traffic volume changes from time to time, a dynamic algorithm is proposed in (Markiewicz et al., 2014). The power consumption minimization problem is formulated using IP model. The objective function jointly minimizes the sum of the power consump-tion of the links and the switches. Four greedy heuristics algorithms are proposed to solve the problem, namely, shortest path first, longest path first, smallest demand first, and highest demand first. The algorithms are tested over two sample topologies, a campus network and mesh network for low, medium and high traffic. Experimental results show that the longest path first gives close to optimal energy saving of up to 35% and is better than the other three algorithms for both topologies.

Traffic aware energy efficiency techniques for cloud computing infrastructure using OpenFlow are presented in (Vu et al., 2014), where Data Manager (DM) is introduced to detect input traffic and control the states of switches. The proposed power manager dynamically controls and updates the operating modes of switches. Based on the information from the DM, the Clock Controller (CC) module changes the frequency of the switch to 0 MHz if the traffic is low. The experimental results show that the approach guarantees QoS. This approach is immune to packet dropping as compared to previous works and results show whole switch saving of about 30–35% on average.

The idea in (Vu et al., 2014) is tested on the NetFPGA platform. Such an approach extends the OpenFlow protocol to include additional information about the frequency each switch is operating. Using this, the controller would be able to allow switches to work at different frequencies. Experimental results show that the proposed switch saves up to 95% when running at low frequency mode. For a large frequency between 2611 mW and 11576 mW, it saves 22.5% energy in total (Vu et al., 2015).

GreenRE (Giroire et al., 2015) uses redundancy removal (RE) to achieve energy efficiency where the motivation is stated as follows. Networks exhibit several redundant links while users access similar contents. Even though redundancy increases reliability, it also degrades the performance of the network. Instead of sending the same data through many different paths repeatedly, sending it through a single link increases the throughput and in effect reduces the load in the links. The links with no loads are subject to turn off. GreenRE presents a work which capitalizes on RE to reduce energy. Experiments conducted on Orange Labs platform demonstrate that RE approach results in 30% less energy consumption. A very important problem addressed in this work is the number of RE enabled routers needed to preserve QoS and reduce energy at the same time. In contrast to the general model we presented, the objective function minimizes only the number of RE-capable routers while minimizing the links.

A bandwidth aware energy efficient technique for data center is presented on the work of (Rui et al., 2015). The controller schedules traffic flows. Traffics can be divided into three statuses, active, queued, and suspended. There is no need for clock synchronization among servers since a centralized controller is responsible for scheduling. A time aware power consumption model of a switch is also presented. A linear programming model is used to capture energy formulation of the entire network. The objective function minimizes the number of active switches and ports which increasing the occupation ratio of ports per switch. Experimental results show that the bandwidth aware technique demonstrates 8.85% higher energy efficiency and lower completion time in Fat-Tree and BCube topologies as relative to the fair share routing.

Energy efficient approaches attempt to minimize the number of power consuming components in the network. The drawback is that QoS may degrade. At the same time there is a need to keep idle network components for fault tolerance and restoration. Restoration and energy efficiency aim at achieving opposite objectives. Restorable Energy Aware Routing with Backup Sharing (REAR-BS) combines both energy efficiency and restoration into a single problem (Xu et al., 2015). A non-linear formulation is used to minimize the number of active links under constraints of maximum utilization and restoration. However, the non-linear formulation is converted to its linear form to get a minimum bound to the optimal solution. An algorithm, Green Restorable Algorithm (GreRA) is proposed to solve the REAR-BS NP-hard problem. Simulation results show that GreRA reduces the energy usage by 15%–50%.

The actual experimentation of energy efficient techniques is costly since it demands operational network. GreenSDN brings an energy efficiency simulation capability on the well known and widely used network emulator Mininet (Lantz et al., 2010) and POX controller (Rodrigues et al., 2015). The simulation tool is tested for both traditional techniques like ALR, switch chip level, switch level, and network-wide scope. The contribution of the study is that it is the first energy efficiency simulation environment built on POX controller. The replication of state-of-the-art approaches has achieved up to 37% energy saving.

An in-band traffic aware energy efficient approach through minimizing the number of active links in multiple controller environment is presented in (Fernandez-Fernandez et al., 2016). The approach presents formal solution for the IP formulated problem and also heuristics. An energy saving of up to 60% is achieved on real world network topology and traffic traces.

The work in (Zhu et al., 2016) integrates energy aware modules in the open source network management platform OpenNaaS. The modules are designed to enable energy monitoring and energy aware routing for different kinds of OpenFlow controllers. Widely known routing algorithms and scheduling algorithms implemented as part of the routing strategies within the framework are evaluated. Simulation results indicate that the priority-based routing leads energy efficiency improvements of 5%–35% compared to other routing strategies, with no degradation in network performance. The contribution of this work lays on adding the scheduler and combining it with known routing algorithms. The experiment is done on 3 different controllers: OpenDaylight, RYU, and Floodlight.

The work in (Riekstin et al., 2016) approaches the energy saving problem through three energy saving capabilities, namely the node component scope, node scope, and network scope. The node components are the hardware components that make up the switch or the host. Examples of node component scope capabilities are the Adaptive Link Rate and Advanced Configurable and Power Interface. These two techniques are Layer 1 energy efficiency capabilities and are also used in the traditional network paradigm. The node scope deals with sleeping the entire switch in response to low traffic. The IEEE 802.3az standard which is the Energy Efficient Ethernet is an example of node scope capability. The network scope energy saving capabilities introduce a more complicated problem that combines the node component and the node scope capabilities. The major problem of applying energy saving mechanism at once is that their effect cancels and the orchestration my come up in a worse energy saving. It might be better to apply one technique alone instead of combining them. This work addresses the effect of the canceling problem by devising a means of consolidating the three capabilities together. Experiments demonstrate energy savings of up to 54% in particular scenarios.

A traffic and resource aware energy saving method is presented in (Rahnamay-Naeini et al., 2016). Unlike most of the related work in energy efficient optimization problems which are linear, the proposed method formulates the problem as non-linear. Instead of considering snapshot of the state of the network to grasp traffic and resources, the system presented introduces the concept of a discrete time. An important question to ask then is how often to pull statistics and solve the optimization problem. The power saving scheme employed is link level. The heuristic algorithm finds the subset of links to be turned off, and it is shown to achieve more than 30% energy saving.

The energy efficiency problem with multiple controllers in the context of 5G network is presented in (Fernández-Fernández et al., 2017). Power consumption is reduced by using traffic engineering techniques, IP formulation for both static and dynamic energy aware routing, and also minimizing the number of active links. As compared to the for-mal solution, the running time of the two heuristics are better but the energy efficiency is sub optimal.

FLOWP (Luo et al., 2017) attempts to achieve both power reduction and QoS for fat-tree topology. An IP formulation for power efficient flow scheduling and the corresponding heuristics is proposed. Unlike the other approaches, the status of network, and the minimum threshold for the utility of links and switches are used in the formulations. Experiments demonstrated an energy saving of 30% and better QoS as compared to ElasticTree and CARPO (Heller et al., 2010; Wang et al., 2012).

The main challenge with energy efficiency techniques is that they exhibit reduced performance. The trade-off between energy saving and performance is simultaneously addressed and IP formulated based on the link utility intervals in our prior works (Assefa and Ozkasap, 2017, 2018). Next Shortest Path and Next Maximum Utility heuristics give priority to performance and energy saving, respectively. However, the framework we presented in (Assefa and Ozkasap, 2018) proposes a single heuristics that achieve both energy saving and performance at the same time by maximizing parameter named the Energy Profit Threshold (EPT). Experiments conducted on real network topology and traces provided by SNDLib show an energy saving of up to 50%.

IP formulation for energy aware routing in carrier-grade Ethernet networks with the objective of turning off links and switches is presented in (Maaloul et al., 2018). A greedy bin packing problem based heuristics named first-fit is presented as a solution. Results show that the first-fit heuristics is scalable and achieves an energy saving of 37%. Unlike other approaches, rule space in the switches is considered as a constraint. The approach is tested using SNDLib real world network topology and traffic traces.

Multiple Topology Switching with Data Plane Forwarding Path Rerouting (MTSDPFPR) is an energy saving approach based on sleeping and rate adaptation of links (Ba et al., 2018). MTSDPFPR is a multi-level dynamic topology switching mechanism thorough finding the maximum utilized link. The approach is tested on real world GEANT network topology and traffic traces. Experimental results show that energy saving of 25% is achieved.

The application scenarios of the traffic aware solutions are generally on campus networks or data centers. However, the specific techniques have switch, port, link, or network wide applicability. The methods that use queue engineering (Nguyen et al., 2013; Wang et al., 2014a; Riekstin et al., 2016; Vu et al., 2015; Chang, 1999) for instance are focused on the switch level queues but their general application is data center or campus networks. The application scenarios in (Heller et al., 2010; Thanh et al., 2012; Wang et al., 2012; Staessens et al., 2011; Giroire et al., 2015; Zhu et al., 2016) are data centers. On the other hand, the methods in (Markiewicz et al., 2014; Fernandez-Fernandez et al., 2016; Assefa and Ozkasap, 2018) are applicable to a general network environments, either specific network topologies or general network topologies.

## 4. End system aware solutions

### 4.1. Overview

A data center consists of interconnected servers (physical hosts) that are structured into racks. The end system connected can be a physical machine or a virtual machine. Server virtualization enables running multiple virtual machines (VM) on a single hardware resource and satisfying the applications' resource demands. Resources include CPU, memory, and network bandwidth. VM migration refers to moving a virtual machine from one host to another for the purpose of achieving energy saving, performance increase, load balancing or system maintenance.

Server consolidation refers to a method used in data centers to minimize the number of active servers by increasing the utility of each server (Ahmad et al., 2015). Hence, instead of operating many servers at low utilization, virtualization technique combines the processing power onto fewer physical servers that operate at a higher total utilization. The deployment of SDN in cloud data center virtual machines boosts QoS and load balancing due to the enhanced flexible control of the network.

Network optimization deals with the energy consumption and communication cost of network components. Focusing only on server consolidation may lead to low network performance. Besides, the network components incur a substantial amount of energy. SDN provides global information about the network, such as the topology, bandwidth utilization, physical machine status, virtual machine status, workload, and other performance statistics. It also provides a flexible way to install forwarding rules so that simultaneous migration of virtual machines can be handled.

Unlike traffic aware techniques, where the network components are the focus for energy saving, end system aware techniques consider both network components and end systems as sources of energy saving. Devising an efficient solution needs periodic updates on the status and organization of the end systems and the traffic at the same time.

### 4.2. End system aware model

The goal of the end system aware solution is to minimize the number of active physical machines through migrating the virtual machines into fewer number of physical machines. Reviewing various models used to solve the problem, we propose a general model for the end system aware energy efficiency. End system aware solutions should be addressed as server consolidation and network optimization problems simultaneously (Jiang et al., 2012; Shirayanagi et al., 2012; Dong et al., 2013). The general model we propose addresses both problems and is inspired by (Jiang et al., 2012).

The problem of virtual machine migration is modeled as a quintuple $(\mathbb{P}, \mathbb{V}, \mathbb{R}p, \mathbb{R}v, \mathbb{R})$ where $\mathbb{P}$, $\mathbb{V}$, $\mathbb{R}p$, $\mathbb{R}v$, and $\mathbb{R}$ correspond to the set of physical machines, set of virtual machines, matrix of resources of physical machines, matrix of resource requirements of virtual machines, and vector of type of resources, respectively. The resources are listed as but not limited to CPU, memory and bandwidth capacities. The dimensions of the matrix $\mathbb{R}p$ and $\mathbb{R}v$ are $|\mathbb{P}|$ by $|\mathbb{R}|$ and $|\mathbb{V}|$ by $|\mathbb{R}|$, respectively.

Let $X$ be a placement matrix where

$$X_{ij} = \begin{cases} 1, & \text{if virtual machine } j \text{ is placed on physical machine } i \\ 0, & \text{otherwise} \end{cases}$$

$P_i^r$ is the amount of resource type $r$ available on physical machine $i$, whereas $V_j^r$ is the demand of virtual machine $j$ for resource type $r$. Let $PM_i$ is a binary variable with value 1 if physical machine $i$ is on, or 0 otherwise.

$$\text{minimize} \sum_i PM_i \tag{8}$$

$$\text{subject to} \sum_{j=1} V_j^r * X_{ij} \leq P_i^r \quad \forall i, r \tag{9}$$

$$\sum_i X_{ij} = 1 \quad \forall j \tag{10}$$

$$PM_i \geq X_{ij} \quad \forall i, j \tag{11}$$

where $1 \leq i \leq |\mathbb{P}|$, $1 \leq j \leq |\mathbb{V}|$ and $r \in \mathbb{R}$.

The first objective function (Eq. (8)) minimizes the number of physical machines turned on. The constraint 9 states the sum of resource demands of virtual machines installed on a given physical machine cannot be more than the capacity of the physical machine. Constraint 10 limits that each virtual machine can be placed on exactly one physical machine. Eq. (11) associates the variables $PM_i$ and $X_{ij}$ by asserting that a physical machine will be turned on or off depending on whether it is used or not.

The second objective that needs to be considered is the network energy consumption. Let $\mathbb{Q}$ be a traffic matrix where $q_{ij}$ is the amount of traffic between $VM_i$ and $VM_j$, and $b_{ij}$ is the number of switches that the traffic between physical machines hosting virtual machines $i$ and $j$ traverses.

**Table 4**
End system aware techniques for energy efficiency.

| Approach | Environment | Network Optimization | Dynamicity | Topology |
| --- | --- | --- | --- | --- |
| Honeyguide (Shirayanagi et al., 2012) | Testbed | ✓ | – | Fat-tree |
| JointVM Placement (Jiang et al., 2012) | Testbed | ✓ | ✓ | Fat-tree |
| Joint Host VM (Jin et al., 2013) | ns-2 | ✓ | ✓ | Fat-tree |
| Joint Host VMP (Zheng et al., 2014) | Testbed | ✓ | – | Fat-tree |
| EQVMP (Wang et al., 2014b) | ns-2 | – | – | – |
| PowerNets (Zheng et al., 2017) | Testbed | ✓ | – | Fat-tree |
| MTAD (Chen et al., 2015) | CloudSim | ✓ | ✓ | – |
| Load Balancing (Carlinet and Perrot, 2016) | Testbed | ✓ | ✓ | – |
| QRVE (Habibi et al., 2016) | Mininet and OpenDaylight | ✓ | – | – |
| NA-VM (Tsygankov and Chen, 2017) | Testbed | ✓ | ✓ | – |
| Int Load Balancing (Hu et al., 2017) | Testbed | ✓ | ✓ | – |
| CloudDC (Liao and Wang, 2018) | Testbed | ✓ | ✓ | Fat-tree |
| SLA-Based (Son et al., 2017) | Testbed | ✓ | ✓ | – |
| Load Balancing (Akbar Neghabi et al., 2019) | Testbed | ✓ | ✓ | – |

$$\text{minimize} \sum_i \sum_j q_{ij} * b_{ij} \quad (12)$$

$$\text{subject to} \sum_{j=1} V_j^r * X_{ij} \le P_i^r \quad \forall i, r \quad (13)$$

$$\sum_i X_{ij} = 1 \quad \forall j \quad (14)$$

The objective function of the general end system aware solution is the combination of the server consolidation and network optimization. Eq. (12) minimizes the communication traffic between virtual machines and the number of switches traversed between physical machines. Eq. (13) constrains the model to respect the sum of the virtual machine resources not to exceed the resources of the physical machine that they are placed on. Eq. (14) asserts that a virtual machine can be placed on exactly one physical machine.

*4.3. Methods*

Table 4 shows the summary of end system aware solutions to energy saving in data centers. The environment column presents the experimental test environment (Testbed, ns-2, or CloudSim) used. The network optimization column shows if an approach considers optimization of the communication cost among virtual machines. Dynamicity is the ability to do server consolidation and network optimization real time. The topology column describes if the method is explicitly applied to a specific topology. The '-' symbol in the table means unknown for the environment and topology columns, or does not apply for the network optimization and dynamicity columns. As shown in Table 4, the majority of the end system aware solutions focus on network optimization in addition to virtual machine placement. Only a few approaches consider online virtual machine migration and traffic consolidation. Unlike traffic aware energy efficient methods, most solutions are topology specific mainly fat-tree topology. Similar to the traffic aware energy efficient techniques in SDN, experiments are conducted on either special testbeds, or simulation and emulation environments.

Virtual machine placement plays a major role in energy savings of data centers. The work in (Kantarci et al., 2012) presents a holistic model for data center power consumption minimization. The model is holistic in the sense it considers all energy incurring components including cooling and virtual machine placement. According to (Dong et al., 2013), a virtual machine placement method should take into account resource constraints such as physical server and network link capacities.

Honeyguide is a virtual machine migration-aware extension of the fat-tree topology network topology for energy efficiency in data center networks (Shirayanagi et al., 2012). Reducing energy consumption is achieved by decreasing the number of active (turned on) networking switches. In this approach, the focus is not only turning off inactive switches, but also trying to maximize the number of inactive switches. To increase the number of inactive switches, two techniques are combined: virtual machine and traffic consolidation. As an extension of existing fat-tree topology, Honeyguide adds bypass links between the upper-tier switches and physical machines. By doing so, it meets the fault tolerance requirement of data centers. Experimental results show relative increase in energy saving of up to 7.8%.

A joint virtual machine placement and routing solution is presented in (Jiang et al., 2012). Dynamic virtual machine placement and routing is modeled as a combinatorial problem. A Markov approximation is used to solve the joint optimization problem under varying workloads. The method used shows better performance as compared to the common heuristic techniques. Unlike other methods ((Shirayanagi et al., 2012; Wang et al., 2014b; Zheng et al., 2014)), this work formulates the workload distribution among virtual machines separately.

An OpenFlow based joint virtual machine placement and flow routing optimization is designed (Jin et al., 2013). Unlike the work of (Jiang et al., 2012) the problem is formulated as an integer linear program and applies a series of techniques to remodel and combine the multi-objective problem into a single objective function. The simulation results show that this approach performs up to 67% energy saving, and it performs better than the existing server consolidation only or network optimization only solutions.

EQVMP proposes energy-efficient and QoS-aware virtual machine placement for SDN based data centers (Wang et al., 2014b). Unlike ElasticTree, power on and off is applied to the servers themselves instead of the switches, the focus is server consolidation. EQVMP combines three techniques: hop reduction, energy saving, and load balancing. Hop reduction divides the VMs into groups and reduces the traffic load among groups by graph partitioning. Energy savings mostly are achieved by VM placement. The motivation behind VM placement is from Best Fit Decreasing and Max-Min Multidimensional Stochastic Bin Packing. Fat-tree is used to represent the VM and servers in the data center. SDN controller is used to balancing the load in the network. Load balancing achieves flow transmission in networks without congestion. Experimental results show an energy saving of 25%.

PowerNets (Zheng et al., 2017) is a joint optimization technique for both server consolidation and network optimization. The peak hours of servers in a data center are not the same. Therefore, the motivation of this work is based on the observation that all the servers do not use their maximum capacity simultaneously. Correlation on the workloads that run on the servers gives a way of grouping servers on their time series. Workloads that use similar servers in a close time interval are grouped to use the same path. By leveraging from server consolidation, DCN optimization and correlation of workloads, PowerNets provides up to 51% energy saving. Different than our general model, Power-Nets takes the correlation of workloads among machines into consideration.

A multiobjective virtual machine placement approximation method is presented in (Chen et al., 2015). The motivation is that virtual machine placement algorithms for cloud data centers find local solutions for each objective and finally combine the results. Such approaches fail to guarantee a global optimum. The multiobjective approximation model hence combines placement efficiency, load balancing, resource utility, and energy efficiency objectives into one. A multi-objective heuristic algorithm for virtual machine placement named MTAD is proposed. MTAD is based on a greedy algorithm, minimum cut and best-fit algorithms. Experiments are conducted on simulation environment and have shown better results than greedy, simulated annealing, and genetic algorithms in the literature. MTAD considers server consolidation, network optimization and also operates online to accommodate dynamic changes.

A load balancing energy-saving approach for geographically distributed data center networks is proposed in (Carlinet and Perrot, 2016). A multi-objective IP formulation that both balances request load and minimizes brown energy consumption by putting hosts to sleep. Experiments conducted using real-world traffic traces exhibit energy savings of up to 42%.

The joint problem of power consumption of data center network, SLA violation, QoS-aware, and VM placement are simultaneously addressed in (Habibi et al., 2016). IP formulation and elephant flow detection based heuristics are presented. Unlike other approaches, a distributed SDN architecture is considered. The experiments are conducted on Mininet emulation environment and OpenDaylight controller achieve 21% energy saving.

A taxonomy for virtual machine placement solutions is presented in (Ortigoza et al., 2016). It also presents metrics to classify virtual machine placement methods by elasticity, overbooking and time awareness. Elasticity is the responsiveness of the system to saturation and under-utility of resources. Server or network overbooking is the re-utilization of resources when they are idle. Since a workload or a service uses a given service for finite length of time, update on the reusability of resources is important.

A variation of multi-commodity minimum cost flow problem is used to model virtual machine migration and load balancing. The two constraints are load balancing and migration time. Whereas the former attempts to achieve virtual machines are evenly distributed to the physical machines, the latter puts a time constraint to do the actual migration. Ant colony heuristics algorithm is provided as a solution. Experimental results show that the migration time is decreased by half as compared to similar approaches. Although this work is not directly related to energy saving, the formulation of the ant colony heuristics can further be studied for efficiency purpose (Tsygankov and Chen, 2017).

A joint optimization effort of energy efficiency and virtual machine migration in a cloud data center is presented in (Liao and Wang, 2018), and a similar approach is also proposed in (Hu et al., 2017). Viewing the virtual machine placement and migration problem as a Multidimensional Vector Bin Packing Problem (MVBPP) is one of the common approaches. However, with the increase in the dimension of the vector, most heuristics fail to find optimal results. In this approach, a mathematical formulation and a hybrid single-parent (partheno-genetic) algorithm to solve the virtual migration integration problem is proposed. Simulation results show that the approach exhibits less migration time as compared to similar approaches. However, it does not fully solve the problem of scalability.

A comparison of several meta-heuristics algorithms used in SDN is presented in (Akbar Neghabi et al., 2019), where different metrics to evaluate meta-heuristics algorithms used for load balancing in SDN are suggested. The metrics are throughput, migration time, energy efficiency, and online migration. Some examples of meta-heuristics algorithms are Genetic algorithm, Ant colony optimization, Particle swarm optimization, Greedy, and Simulated annealing. The major reason to resort to such algorithms is due to the fact that the formal solution is not time efficient and converges for a small number of physical and virtual machines. On the other hand, the meta-heuristics algorithms provide sub-optimal solutions and in some cases, they are far from optimal.

A Service Level Agreement (SLA) aware resource overbooking for energy efficient workload allocation and virtual machine placement is presented in (Son et al., 2017). The approach solves the fixed amount of resource allocation using SDN to dynamically consolidate traffic and control of Quality of Service (QoS) from a logically central point. It jointly minimizes the power consumption of the host and the network and the number of SLA violations.

The application scenarios of the end system aware solutions are SDN based data centers, either centralized or geographically distributed. The energy saving objective focuses on minimizing the number of active physical machines through migrating virtual machines. Moreover, balancing the workload on virtual machines is also another focus that needs to be considered. The end system aware methods we presented in this work are applicable to scenarios where there are virtual machines and workloads that can be assigned and migrated to different machines.

## 5. Rule placement solutions

### 5.1. Overview

Routing in SDN is determined by the control plane and applied by forwarding in the data plane. For each new flow, its first packet triggers a flow initiation request from a switch to the controller. Then, the path for the packets of the flow is determined by the controller and the corresponding flow table entries are installed on the switches of the path. Rules are generated and placed proactively or reactively. In proactive rule placement, the controller generates the rules and pushes to respective switches initially. This approach demands network requirements and policies to be known beforehand. In reactive rule placement, rules are generated in response to changes in the network policy, topology, status of switches or traffic (Kang et al., 2013; Kanizo et al., 2013; Giroire et al., 2014).

An end-point policy is a network service requirement which decides the outgoing link that each packet should be forwarded. On the other hand, a routing policy describes the actual path for each packet. Rule space optimization can be applied locally at switch level to increase efficiency in handling installed rules, or can be viewed as a network-level problem that generates a set of rules, compresses them and pushes them into the forwarding switches. This section presents global network-wide placement on SDN pertaining to energy efficiency.

The energy efficiency techniques in the context of rule placement solutions develop the energy cost model with the constraints, then propose heuristic algorithms that provide optimum energy saving. The controller generates the forwarding rules and sends them to the switches. Placing the rules to respective switches distributed across the network and optimizing an objective function under the constraints is NP-hard problem. The objective function in our particular case is minimizing energy consumption whereas the constraints are the number of switches, flow table capacity, link capacity, and number of ports per switch.

### 5.2. Rule placement model

Based on the review of models on rule placement in SDN, we propose a simplified and general model for optimization of energy. Energy efficient strategies can be achieved by memory minimization or number of links used or the combination of the two. Our approach is inspired by (Nguyen et al., 2015) and (Nguyen et al., 2014), and in our model the objective is to minimize the total number of rules installed in the switches. The general model can be extended if there is a need to add different constraints or parameters.

**Table 5**
Rule placement techniques for energy efficiency.

| Approach | Rule Meaning | Compromised Routing | Rule Space Reduction | Proactive | Reactive | Wildcard |
| --- | --- | --- | --- | --- | --- | --- |
| Big Switch (Kang et al., 2013) | – | – | – | ✓ | ✓ | – |
| Palette (Kanizo et al., 2013) | – | – | – | ✓ | – | – |
| Optimizing rule placement (Giroire et al., 2014) | ✓ | ✓ | – | ✓ | – | ✓ |
| Two-dimensional compression (Giroire et al., 2016) | ✓ | – | – | ✓ | – | ✓ |
| OFFICER (Nguyen et al., 2015) | ✓ | ✓ | ✓ | ✓ | – | – |
| MINNIE (Rifai et al., 2015) | – | – | ✓ | ✓ | – | ✓ |
| MINNIE-Extended (Rifai et al., 2017) | – | – | ✓ | ✓ | ✓ | – |
| Flow Aggregation (Kosugiyama et al., 2017) | – | – | ✓ | ✓ | ✓ | ✓ |
| Wildcard (Huin, 2016) | – | ✓ | ✓ | ✓ | – | ✓ |
| MIRA-RA (Ashraf, 2017) | – | – | ✓ | – | ✓ | – |
| FTRS (Leng et al., 2017) | – | – | ✓ | ✓ | – | – |
| Reduce Flow configuration (Galán-Jiménez et al., 2018) | – | – | ✓ | ✓ | – | – |

The network is represented as a directed graph $\mathbb{G}=(\mathbb{Z},\mathbb{E})$ where $\mathbb{Z}$ is the set of switches and $Z_i \in \mathbb{Z}$ represents switch $i$ and $e_{ij} \in \mathbb{E}$ represents that there exists a link between switches $Z_i$ and $Z_j$. $W_{ij}$ is the bandwidth of the link connecting switches $Z_i$ and $Z_j$. Let binary variable $S_i$ denote the status of switch $Z_i$ as stated in traffic aware model. $G_i$ is the maximum number of rules that can be installed in switch $Z_i$, respectively. There are two types of external nodes $H_{in}$ and $H_{eg}$ which are ingress hosts and egress hosts, respectively. $\lambda_f$ is the packet rate of flow $f \in \mathbb{F}$.

$\mathbb{A}$ is an allocation matrix of rules representing flows to switches, where

$$a_{if} = \begin{cases} 1, & \text{if rule representing flow } f \text{ is installed on switch } Z_i \\ 0, & \text{otherwise} \end{cases}$$

The dimension of the allocation matrix is the number of switches by the number of flows.

$L_{ij}$ a binary variable indicating whether edge $e_{ij}$ is in active state or not.

$$L_{ij} = \begin{cases} 1, & \text{if edge } e_{ij} \text{ is active} \\ 0, & \text{otherwise} \end{cases}$$

$K_{ij}$ is a binary variable showing if at least one flow is passing through the edge $e_{ij}$

$$K_{ij} = \begin{cases} 1, & \sum_{f \in F} F_{ij} \geq 1 \\ 0, & \text{otherwise} \end{cases}$$

$$\text{minimize } \sum a_{if} \; \forall f \in \mathbb{F} \tag{15}$$

$$\text{subject to } \sum \lambda_f * F_{ij} \leq W_{ij}, \forall e_{ij} \tag{16}$$

$$\sum_{f \in \mathbb{F}} F_{ij} \leq G_i, \forall Z_i \tag{17}$$

$$a_{if} = F_{ij} \text{ and } a_{jf} = F_{ij}, \forall F_{ij} \tag{18}$$

$$L_{ij} \leq \sum_{f \in \mathbb{F}} F_{ij}, \forall e_{ij} \tag{19}$$

$$L_{ij} \geq K_{ij}, \forall e_{ij} \tag{20}$$

$$\sum_{\forall f \in \mathbb{F}} F_{iu} = \sum_{\forall f} F_{vj} Z_i \text{ is connected to } H_{in}$$

and $Z_j$ is connected to $H_{eg}$ (21)

The objective function in Eq. (15) minimizes the number of rules installed in the switches. The constraint 16 states that the total rate of flows on a given link cannot be more than the link capacity. The constraint 17 indicates that the number of rules installed in a switch cannot be more than the capacity of the switch. Eq. (18) states that if a flow passes through edge $e_{ij}$, its corresponding rule should be installed on the switches $Z_i$ and $Z_j$. Eqs. (19) and (20) maintain the relation between flow $f$ and link state $L_{ij}$ parameters, where Eq. (19) forces links to be put into inactive state if no flow is passing through them, Eq. (20) asserts the reverse by forcing a link to be put into active state if it is used by one or more flows. Constraint in Eq. (21) is about conservation of flow which indicates the number of flows entering and leaving a network should be equal.

In addition to the constraints stated above, the model can be tailored to include topology awareness and time changes. When the model is complex, a feasible solution can be found at the cost of performance. Relaxation on the other hand gives better performance but it only provides the boundary of the solution.

### 5.3. Methods

Table 5 presents the summary of rule-placement techniques used to optimize energy efficiency. All the rule placement methods assume a global view of the network. Some approaches compromise the routing policy. Compromised routing however, comes with extra cost of using longer routes. The end-point policy needs to be respected unconditionally. Approaches also focus on reducing the rule space. Reduced rule space can lead to fewer number of switches and hence reduce the energy cost. Table 5 focuses on the following features. The meaning of the rules is defined as what further information can be found out from the rules. A good example can be packet types, or similar pattern of source or destination addresses. For a performance demanding network environment, every flow tends to choose the shortest route. For energy saving however, switches can be turned off or links can be made inactive. This forces the flows to choose alternative paths which can be longer. The compromise in routing saves energy but at a cost of longer paths. Rule space reduction is the attempt of minimizing the global number of rules using various techniques. Proactive and reactive rule placements are rule allocation techniques prior to packet arrival and in response to packet arrival, respectively. Wildcard column shows approaches that use wildcards to compress flow rules.

Big Switch approach utilizes the fact that SDN controllers have a global view of the network and proposes that the entire network should be viewed as one big switch (Kang et al., 2013). The architecture considers the SDN controller with three components: end-point policy, routing policy and rule placement policy. A high-level SDN application defines end-point connectivity policy on top of big switch abstraction; a mid-level SDN infrastructure layer should decide on the routing policy; and a compiler should synthesize on the end-point and routing policy and develop an effective set of forwarding rules that obey the user-defined policies and adhere to the resource constraints of the underlying hardware. Minimizing the number of rules needed to real-ize the end-point policy under rule capacity constraint is both a decision and an optimization problem. The architecture addresses the two problems through a heuristic algorithm that recursively covers the rules and packs groups of rules into switches along the path.

The Palette distribution framework is a distributed approach applied to SDN tables (Kanizo et al., 2013). Since the SDN controller table can only handle hundreds of entries, and the memory is expensive and power hungry, Palette decomposes large SDN tables into small ones and then distributes them across the network, while preserving the overall SDN policy semantics. Palette helps balance the sizes of the tables across the network, as well as reduces the total number of entries by sharing resources among different connections. It copes with two NP-hard optimization problems: decomposing a large SDN table into equivalent sub-tables, and distributing the sub-tables. The problem of traversing is formulated using the rainbow problem. By giving unique color for each sub-tree, each connection traverses each color type at least once. Implementation of Palette framework is based on graph theory formulation algorithms and heuristics.

The drawbacks of Palette and Big Switch approaches are that they do not rely on the meaning of the rules. The main objective in both cases is to push rules based on the path of the flow which means path determines the rules. The meaning of the rules however helps to aggregate rules for instance rules which have the similar pattern of source and destination addresses, or rules whose type of packet is TCP or UDP. A technique of compacting rules, which optimizes the rule placement, is proposed in (Giroire et al., 2014). This approach analyzes the meaning of the rules, together with heuristic optimization method, to minimize energy consumption for a backbone network while respecting capacity constraints on links and rule space constraints on routers. Relaxation on the routing policy is proven to yield better efficiency but can result in longer paths. They present an exact formulation using integer linear programming, and introduce efficient greedy heuristic algorithm for large networks.

Methods using wildcard rules are designed to reduce the rule space for each switch and also to reduce the communication overhead between switches and the controller. The work in (Giroire et al., 2016) presents a theoretical analysis of compressing rules and states that the problem is NP-Hard. It also presents results of NP-completeness of fixed-parameter tractability and approximation algorithms.

OFFICER is a general framework that models the network as black box respecting end-point policy strictly but relaxes the routing policy (Nguyen et al., 2015). The idea of relaxation in routing policy for better efficiency is motivated by the work in (Nguyen et al., 2014). But the relaxation of routing policy can cause the drawback of longer paths. OFFICER framework proposes an optimization model and three types of heuristics.

Another rule space reduction solution to the rule placement problem is MINNIE (Rifai et al., 2015), which is an algorithm that aims at computing load-balanced routes. It models the network as a directed graph and provides a two phase solution: the compression phase and the routing phase. The compression phase proposes a heuristic solution which is motivated by the proof that the rule compression using wildcard is NP-Hard problem (Giroire et al., 2016). The routing phase also provides a heuristic based on shortest path algorithm with adaptive metric. Experiments show that the smart rule allocation reduces the number of rules needed by 50%. The scalability issue of MINNIE is addressed in (Rifai et al., 2017), and it is extended to work online in a dynamic environment. Moreover, the experiments are conducted on different network topologies in addition to the fat-tree. The compression of the rules is done in response to every new flow arrival. Experimental results show that the extended approach performs better in terms of reduced packet loss rate and delay.

Flow aggregation is one method where the flow entries are aggregated per switch based on their destination or other criteria so that all can egress to similar port (Kosugiyama et al., 2017). In this work, IP formulation of the flow aggregation problem is presented. The heuristics proposed first aggregates the flow that is already installed, and then the aggregation is done for each new arriving flow. Unlike other aggregation and compression techniques, this approach considers end-to-end routing policies.

Two ILP formulations for flow table compression using wildcard rule are presented in (Huin, 2016). The first ILP formulation is based on default port compression where the main objective is to minimize the number of active links by optimizing the number of flows passing through the default port. The default port is a special port in each switch, from which unmatched packets are sent to the controller. This formulation considers a single field meaning that wildcard compression is applied to either source or destination information of the table but not both. The second ILP formulation extends the first one by allowing multiple field compression. However, the energy saving results of the two are similar.

Minimum Rule Application (MIRA) proposes an IP formulation that minimizes the number of rules installed in dynamic traffic (Ashraf, 2017). As traffic evolves in the network, this forces online rule placement algorithms to generate undesired traffic between switches and the controller to modify flow entries. To overcome the NP-harness of the problem, a rule aggregation heuristics named MIRA-RA is proposed. Their approach attempts to jointly reduce the number of links and also increase the utility of under-utilized links. First Table Reduction Scheme (FTRS) on the other hand reduces the number of exited flow entries by combining the entries that are less important into fewer new ones. Experiments conducted on star topology, tree topology and distributed topology shows that FTRS is able to reduce the number of flow entries by 99%.

The work in (Galán-Jiménez et al., 2018) is an attempt to reduce the re-configuration cost on flow tables when traffic demand changes. The problem is formulated using IP and unlike other approaches, it considers the obsolete flow entries that need to be removed and the new flow rules that need to be installed. A Genetic Algorithm based heuristics is proposed to simultaneously minimize the network power consumption and the number of modified rules in flow tables. Experimental results conducted on SNDlib (Orlowski et al., 2007) network topology and traces show that up to 100 times rule size reduction and more than 20% power reduction is achieved as compared to similar approaches.

The application scenarios of the rule placement based energy saving techniques are basically the flow tables in the switches. The energy efficiency formulations and heuristics are realized through pushing the rules. However, the rule space optimization is observed from global view of the network. Such solutions can be useful in network environments that have a capability of compressing rules.

## 6. Hardware-based solutions

### 6.1. Overview

The three groups of techniques presented, namely traffic aware, end system aware, and rule placement, are all software-based. In this section, we discuss the hardware advancements needed to make SDN enabled switches power efficient. Such approaches attempt to minimize the memory need of information stored in forwarding switches. In SDN, forwarding switches use Ternary Content Addressable Memory (TCAM), which is a specialized type of high-speed memory that performs an entire memory search in a single clock cycle.

Content Addressable Memory (CAM) is a special type of memory that enables direct query to the content without having to specify its address. A search in CAM provides a search tag which is the representation or the content itself, and returns the address of the content if the item is found. The content is represented in binary. TCAM is a specialized CAM, and the term ternary refers to the memory's ability to store and query data using three different inputs: 0, 1 and X. The X input, which is often referred to as a wild-card state, enables TCAM to perform broader searches based on pattern matching. TCAM is popular in SDN switches for fast routing lookup and is much faster than the RAM. However, TCAM is expensive, consumes high amount of power, and generates a high level of heat. For example, a 1 Mb TCAM chip consumes 15–30 W of power, and TCAM is at least as power hungry as SDRAM. Power consumption together with the consequent heat generation is a serious problem for core routers and other networking devices (Applegate et al., 2007; Meiners et al., 2007, 2012; Kannan and Banerjee, 2012; Banerjee and Kannan, 2014).

**Table 6**
Hardware-based techniques.

| Approach | Geometric Model | Path | Flow | Tag |
|---|---|---|---|---|
| Rectilinear (Applegate et al., 2007) | ✓ | – | ✓ | ✓ |
| TCAM Razor (Meiners et al., 2007) | ✓ | – | ✓ | – |
| Bit Weaving (Meiners et al., 2012) | – | – | ✓ | ✓ |
| Compact TCAM (Kannan and Banerjee, 2012) | – | – | ✓ | ✓ |
| Tag-in-Tag (Banerjee and Kannan, 2014) | – | ✓ | ✓ | ✓ |

Two kinds of compression that are applied in TCAM are rule compression and content compression. In traditional access control list, a rule has five components: source range, destination range, protocol, port(s), and action. In SDN, the forwarding decision of a switch is based on flow tables implemented in TCAM. Each entry in the flow table defines a matching rule and is associated with an action. Upon receiving a packet, a switch identifies the highest-priority rule with a matching predicate, and performs the corresponding action. The proposals in (Applegate et al., 2007; Meiners et al., 2007, 2012; Kannan and Banerjee, 2012) are attempts to compact these rules to utilize TCAM effectively.

### 6.2. Methods

Table 6 gives examples of hardware-based techniques to enhance energy efficiency in a switch. The hardware-based solutions differ based on the type data of presentation used in the TCAM and what information is used to describe a the flow table entry. Geometric models such as graphs are used in the representation of the flow entries. (Applegate et al., 2007) and (Meiners et al., 2007) use geometric data structure to represent the content in the TCAM. Path column describes in some approaches the path of the flow from source to destination is stored in the TCAM at the cost of memory. Flow column describes if the whole flow information is stored in the memory. Tag representation considerably decreases the size of information stored in the TCAM as compared to path and flow.

Rectilinear (Applegate et al., 2007) is an approach that exploits SDN features such as programming interface to the switches and dynamic determination of actions for each flow at the switches. The compacting reduces the size of bits to store information that are essential to classify packets to a flow. A flow-id is given to each flow to uniquely identify packets in the corresponding flow. The packet headers are modified at the forwarding switches to carry the flow-id that will used by other switches on the path for classifying the packets. A shorter tag representation for identifying flows than the original number of bits are used to store the flow entries of SDN switches. The authors demonstrated that the compact representation on flow can reduce 80% TCAM power consumption on average.

TCAM Razor proposes a four step solution to compress the packet classifier (Meiners et al., 2007). First, it converts a given packet classifier to a reduced decision diagram. Second, for every non-terminal node in the decision diagram, it minimizes the number of prefixes associated with its outgoing edges using dynamic programming. Third, it generates rules from the decision diagram. Last, it removes redundant rules.

The Bit Weaving technique employs a non-prefix compression scheme (Meiners et al., 2012), and is based on the observation that TCAM entries that have the same decision but whose predicates differ by only one bit are merged into one entry by replacing the bit in question with *. Bit weaving consists of two new techniques, bit swapping and bit merging. First it swaps bits to make them similar and then merges such rules together. The key advantages of bit weaving are that it runs fast, it is effective, and can be complementary to other TCAM optimization methods as a pre/post-processing routine.

A Compact TCAM approach that reduces the size of the flow entries is proposed in (Kannan and Banerjee, 2012) that uses shorter tags for identifying flows than the original number of bits used to store the flow entries for SDN switches. The catch for this approach comes from the dynamic programming capability of SDN to route the packets using these tags. Furthermore, the usage of SDN framework to optimize the TCAM space is introduced.

Unlike the other approaches, Tag-in-Tag technique (Banerjee and Kannan, 2014) does not apply optimization on the flow table entries directly. It rather converts the flow table entries into a structure of Path Tag (PT) and Flow Tag (FT). PT is used for routing the packets and FT is used to correlate each packet with its flow. By encapsulating the FT with PT, the approach utilizes path similarity and also provides a fine grained management of resources. Experiments using both real world and artificial data, and analysis results show that Tag-in-Tag technique reduces per-flow power consumption by 80% and also reduces the size of the flow entries 15 times on average.

## 7. Open issues and discussion

In this section, we indicate and discuss the open issues and future research directions for energy efficiency in SDN. Our survey indicates that most of the recent approaches for energy efficiency in SDN reside in the category of traffic aware solutions. However, there exist areas for improvement and open issues that can be addressed.

### 7.1. Open issues in software based solutions

In the traffic aware approaches, turning the network components on and off based on traffic characteristics and link utilization helps in reducing energy consumption. However, determining the set of network components to turn on or turn off dynamically without affecting QoS and network performance is an NP-hard problem. An efficient solution in this area should consider the trade-off between energy savings and network performance. The general optimization model we presented in this survey can be tailored to consider objectives of network performance in addition to energy efficiency.

#### 7.1.1. Traffic aware
Scalability, flexibility and security are the other open research issues in traffic aware energy efficiency. The traffic aware methods should support scalability as the traffic load gets larger. Another dimension of scalability can be considered for multi-domain SDN scenarios where each network is managed by a controller and there exist a need for a scalable solution for energy efficiency across multiple controllers. Extending an energy efficient solution for a single domain to multiple domains would not be straightforward, and needs separate optimization models. Flexibility that is the ability of the system to adapt dynamic network conditions such as different topologies, updates in the number of nodes, failures in network components, should be considered. Solutions addressing energy efficiency and meeting requirements of secure communications among controllers and switches as well as among controllers in multi-domain settings would also be valuable.

*7.1.2. End system aware*

End system aware approaches are based on the idea of virtual machines. Different than traffic aware techniques, where the network components are the focus for energy saving, end system aware techniques consider both network components and end systems as sources of energy saving. Devising an efficient solution needs periodic updates on the status and organization of the end systems and the traffic at the same time. On the other hand, end system awareness considers two types of problems that are server consolidation and network optimization, often addressed separately. Having multiple-objective function however is computationally expensive where formal methods do not scale with the size of the network. In these directions, improved formulation and efficient heuristics need to be devised. The general optimization model we proposed for end system aware approach can be used a baseline for this purpose.

*7.1.3. Rule placement*

A rule placement mechanism directly affects the network performance and also determines the routing. Forwarding rules are determined and pushed to the switches by the controller. Thus, rule space optimization in SDN can be formulated as a global network-level problem. Placing the rules to respective switches distributed across the network and optimizing an objective function under the constraints is NP-hard problem. Given a routing policy and end-point policy of the network, solutions for space efficient rule representation that would lead to energy savings are necessary. Energy efficient solutions in this group need to formalize the energy cost model and the constraints associated, then apply heuristic algorithms to find optimum energy saving strategy. In this context, the general optimization model we presented for rule placement solutions would provide guidelines for the objective functions, parameters and constraints to be considered.

*7.2. Open issues in hardware based solutions*

In the context of hardware based solutions, since SDN switches use TCAM to keep flow tables, which is expensive and power hungry memory, optimizing the memory space allocated to flow tables would help in reducing overall energy consumption. For example, some solutions aim at compacting the rules and hence reducing memory space requirements. However, for such compacting TCAM solutions, information stored in TCAM cannot be further compressed beyond a certain threshold. Optimizations focusing on memory space reduction by considering minimum energy consumption and reduced flow table update rates would be valuable. Besides, traffic aware solutions could incorporate energy efficiency component in terms of memory usage of switches.

In general, designing a unified framework for evaluation and comparison of software based energy efficiency solutions in SDN would be valuable. Formal definition of the energy saving problem in SDN is the base for applying a sound theoretical solution. The general optimization models presented in this article would be helpful for research studies in these directions. In an energy efficient controller, designing modules for traffic awareness, end system awareness, and rule placement that operate in a dynamic environment while maintaining quality of service is another open issue to explore.

## 8. Conclusion

Energy awareness has become an important design requirement for modern networking mechanisms, and designing energy efficient solutions is non-trivial since they need to tackle trade-off between energy efficiency and network performance such as QoS requirements and dynamic network conditions. In this article, we address the energy efficiency capabilities that can be utilized in the emerging software defined networks. We provide a comprehensive and novel classification of software-based energy efficient solutions in SDN into subcategories of traffic aware, end system aware and rule placement. For the recent developments in the area, we identify the key features of solutions and system designs in each category. Furthermore, we propose general optimization models for each subcategory, and present the objective function, the parameters to be considered and constraints that need to be respected for each model. Detailed information on the characteristics of state-of-the-art solutions for each category, their advantages, drawbacks are provided. Hardware-based solutions used to enhance the efficiency of switches are also discussed.

Despite the fact that there exist some solutions available for different dimensions of energy efficiency in SDN, there are open research issues, challenges and areas for improvement. Thus, we also discussed open issues and research directions in this area. Through this article, the reader can gain comprehensive knowledge on the state-of-the-art methods for energy efficiency in SDN, key properties, classification and general optimization model for each software-based category. The general optimization models proposed for each category would provide guidelines for the parameters and constraints to be considered in future studies.

## Acknowledgments

This work was partially supported by the COST (European Cooperation in Science and Technology) framework, under Action IC0804 (Energy Efficiency in Large Scale Distributed Systems), and by TUBITAK (The Scientific and Technical Research Council of Turkey) under Grant 109M761. A very preliminary version of this work was presented in ICN conference (Assefa and Ozkasap, 2015).

**Beakal Gizachew Assefa** is a Ph.D. Candidate, teaching and research assistant in the Department of Computer Engineering at Koç University. He received his MS in Computer Engineering from Izmir Institute of Technology, Izmir, Turkey in 2012. He received his BSc degree in Computer Science and Information Technology from Haramaya University, Haramaya Ethiopia. His research interests include databases, software-defined networks, energy efficiency, machine learning, and distributed systems.

**Öznur Özkasap** received the M.S., and Ph.D. degrees in computer engineering from Ege University, Izmir, Turkey, in 1994, and 2000, respectively. From 1997 to 1999, she was a Graduate Research Assistant with the Department of Computer Science, Cornell University, Ithaca, NY, USA, where she completed her Ph.D. dissertation. She is currently a Professor with the Department of Computer Engineering, Koç University, Istanbul, Turkey, which she joined in 2000. Her research interests include distributed systems, multicast protocols, peer-to-peer systems, bio-inspired distributed algorithms, mobile and vehicular ad hoc networks, energy efficiency, cloud computing, and computer networks. Dr. Özkasap serves as an Area Editor of the Future Generation Computer Systems journal, Elsevier Science, and National Representative of ACM-W Europe. She also served as an Area Editor of the Computer Networks journal, Elsevier Science, as a Management Committee Member of the European COST Action IC0804: Energy efficiency in large-scale distributed systems, and as a member of the European COST Action 279: Analysis and Design of Advanced Multiservice Networks supporting Mobility, Multimedia, and Internetworking. She is a recipient of the Turk Telekom Collaborative Research Awards in 2012, the Career Award of TUBITAK (The Scientific and Technological Research Council of Turkey) in 2004, and TUBITAK/NATO A2 Ph.D. Research Scholarship Abroad in 1997, and she was awarded Teaching Innovation Grants by Koç University.